\documentclass[showpacs,aps,prb,twocolumn,groupedaddress]{revtex4}

\usepackage{graphicx}

\begin{document}


\title{Shubnikov - de Haas effect in the quantum vortex liquid state of the organic superconductor $\kappa$-(BEDT-TTF)$_{2}$Cu(NCS)$_{2}$}


\author{T. Sasaki}
\author{T. Fukuda}
\author{N. Yoneyama}
\author{N. Kobayashi}
\affiliation{Institute for Materials Research, Tohoku University, Katahira 2-1-1, Sendai 980-8577, Japan}


\date{\today}

\begin{abstract}
We report the Shubnikov-de Haas (SdH) oscillations observed in the vortex liquid state of the quasi two dimensional organic superconductor $\kappa$-(BEDT-TTF)$_{2}$Cu(NCS)$_{2}$.  
The SdH oscillations can be observed down to about 5~T at 0.5~K, where the flux flow resistivity becomes as small as about 30 \% of the normal state value.  
Below the upper critical field $H_{\rm c2}$ of about 7~T, the additional damping of the SdH oscillation amplitude appears, as well as that of the de Haas-van Alphen (dHvA) oscillations, with respect to the normal state one which is described with the standard Lifshitz-Kosevich formula.  
The magnitude of the additional damping near $H_{\rm c2}$ is the same with that observed in the dHvA oscillations and well explained by the theoretical predictions in consideration of fluctuations in the thermal vortex liquid state.    
In the quantum fluctuation region at lower temperature, however, only SdH effect shows the stronger damping than that of the dHvA oscillations. 
The different magnetic field dependence of the additional damping of the oscillation amplitude between the SdH and dHvA effects is discussed in connection with the effect of the transport current on the short-range order of vortices in the quantum vortex slush state reported at the same temperature and magnetic field region.

\end{abstract}



\maketitle


\section{Introduction}

After the report of the magnetic quantum oscillations in the superconducting state of 2{\it H}-NbSe$_{2}$ more than quarter century ago,\cite{graebner} the oscillations of the magnetization, de Haas - van Alphen (dHvA) effect, in the vortex state seems to be confirmed experimentally for a variety of the type-II superconductors in the last decade.\cite{janssen}  
Common experimental results of the dHvA oscillations in the superconducting state are that the additional damping of the oscillation amplitude appears below the upper critical field $H_{\rm c2}$ with respect to the normal state damping.  
The additional damping has been discussed in several ways theoretically.\cite{maniv1}  
The questions in those discussions are summarized as follows;\cite{maniv1}
(1) if the superconducting gap $\Delta_{0}$ exists in the vortex state just below $H_{\rm c2}$, it would have drastically damped the oscillations at low temperatures by a factor of $\exp{(-\Delta_{0}/k_{B}T)}$, 
(2) the inhomogeneous field distribution due to the flux lattice would broaden the Landau levels, 
and (3) the inhomogeneity in the superconducting order parameter associated with the vortex lattice leads to inhomogeneous broadening of the Landau levels in the quasiparticle spectrum near the Fermi surface.  
Thus the oscillations include basically rich information on the quasiparticle in magnetic fields and also the vortex matter properties.\cite{blatter}
  
Shubnikov-de Haas (SdH) oscillations in the superconducting state is very difficult to be observed because the finite resistivity is needed inevitably.  
Then it can appear in the quite limited field-temperature region where the long-range translational order of the vortices is lost.  
Quasi two dimensional (Q2D) organic superconductors are good candidate for the observation of SdH effect in the vortex state because large fluctuations induce the wide vortex liquid region.\cite{lang1,friemel}  
At low temperature below 1 K, the quantum vortex liquid (QVL) is realized in $\kappa$-(BEDT-TTF)$_{2}$Cu(NCS)$_{2}$ ($T_{\rm c} \simeq$ 10 K) due to the large quantum fluctuation instead of the thermal one.\cite{sasaki1,mola}  
In the QVL region, the finite resistivity is expected to remain even below $H_{\rm c2}(T \simeq 0)$.  
Recently the transport properties in the QVL region have been examined in detail, and the finite resistivity has been confirmed.\cite{sasaki2}

In this paper, we report the SdH oscillations observed in the QVL region of the Q2D organic superconductor $\kappa$-(BEDT-TTF)$_{2}$Cu(NCS)$_{2}$.  
An additional damping of the oscillation amplitude appears on both the SdH and dHvA effects around $H_{\rm c2}$, which may come from the superconducting fluctuation.  
At lower temperature in the QVL region, however, the stronger damping is observed only on the SdH effect.  
The different magnetic field dependence of the additional damping of the oscillation amplitude between the SdH and dHvA effects is discussed in connection with the effect of the transport current on the short-range order of vortices in the quantum vortex slush state\cite{sasaki2} reported at the same temperature and magnetic field region.

\section{Experiment}

High quality single crystals of $\kappa$-(BEDT-TTF)$_{2}$Cu(NCS)$_{2}$ were grown by an electrochemical oxidation method.  
The magnetic torque measurements were performed by using precision capacitance torquemeter.  
The in-plane and the out of plane resistivities were measured along the $b$ and $a^{*}$ axes, respectively, by means of a conventional ac or dc four terminal method.  
The electrical terminals were made of evaporated gold films, and gold wires (10 ${\mu}$m) were glued onto the films with gold or silver paint.  
The contact resistance was about 10 $\Omega$ for each contact at room temperature, but it became less than 1 $\Omega$ at low temperature where the experiments were carried out.
The torquemeter and the samples for the resistivity measurements were fixed to the single axis rotation holder which can change the sample direction with respect to the magnetic field with the accuracy of 0.05 degree.  
The holder with the samples was cooled slowly from room temperature to 4.2~K in 48 hours and specially slow cooling rate was used between 50 and 75~K in order to avoid the disorder of the terminal ethylene group of the BEDT-TTF molecules. \cite{mueller1} 
The holder was directly immersed in liquid $^{3}$He of the refrigerator which was combined with a 15 T superconducting magnet at the High Magnetic Field Laboratory for Superconducting Materials (HFLSM), IMR, Tohoku University. 
The results presented in this paper were obtained on three samples \#1 for the magnetic torque, and \#2 and \#3 for the resistivity measurements from different batches.  
We found that other one sample for the magnetic torque and one sample for the resistivity gave qualitatively similar results which were not presented in this paper.

\section{Results and discussion}

\begin{figure}[t]
\includegraphics[viewport=1cm 4cm 21cm 28cm,clip,width=0.9\linewidth]{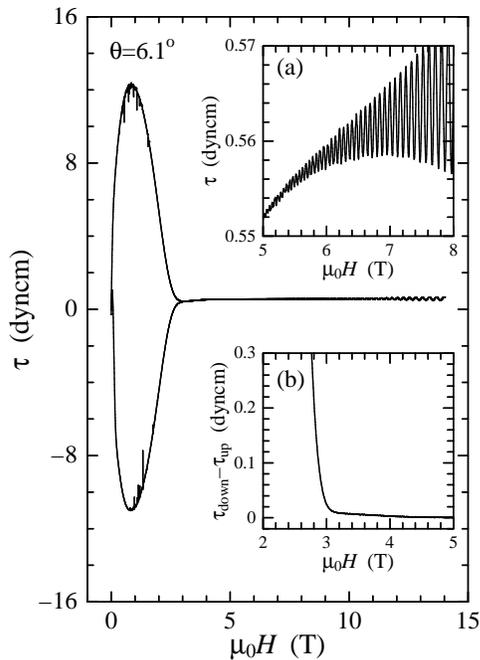}
\caption{Magnetic torque curves of the sample \#1 in $\kappa$-(BEDT-TTF)$_{2}$Cu(NCS)$_{2}$ at 0.52 K and in the field direction tilted by $\theta =$6.1$^{\circ}$ from the $a^{*}$ axis.  (a) the dHvA oscillations around 7~T in the expanded scale. (b) the magnitude of the torque hysteresis. }
\end{figure}

Figure 1 shows the magnetic torque curves of the sample \#1 at $T =$ 0.52~K.  
The overall features are the same with the previous report;\cite{sasaki1} the irreversible and reversible regions are separated at $H_{\rm irr} \sim$ 3~T (Fig. 1(b)), and the dHvA oscillations with one fundamental frequency of $F_{\alpha} =$ 599 $\pm$ 2~T are observed in both the normal and superconducting states.  
The quality of the sample used in this study seems to be better than that in the previous work\cite{sasaki1} judging from the large amplitude of the dHvA oscillations.  
The reversible magnetic torque region ($H_{\rm irr} \simeq$ 3~T $< H <$ $H_{\rm c2} \simeq$ 7~T) below 1~K is expected to be the QVL region.  
The finite resistivity appears in the QVL region even at $T \sim$ 0. 
The detail of the transport properties in the QVL region has been already reported.\cite{sasaki2}  
It is noted only here that a weak non-linear behavior of the resistivity is found in the QVL region.  
Such non-linearity is not observed in the thermal vortex liquid (TVL) region above 1 K.  
The concept of the quantum vortex slush has been proposed for the non-linear behavior below 1~K.  
The vortex slush with only the short-range order of vortices has been found in the high-$T_{\rm c}$ oxides.\cite{worthington,nishizaki,nonomura}  
The effect of the quantum vortex slush and the TVL states on the dHvA and SdH effects will be discussed latter. 

\begin{figure}[t]
\includegraphics[viewport=1cm 6.5cm 21cm 24cm,clip,width=0.9\linewidth]{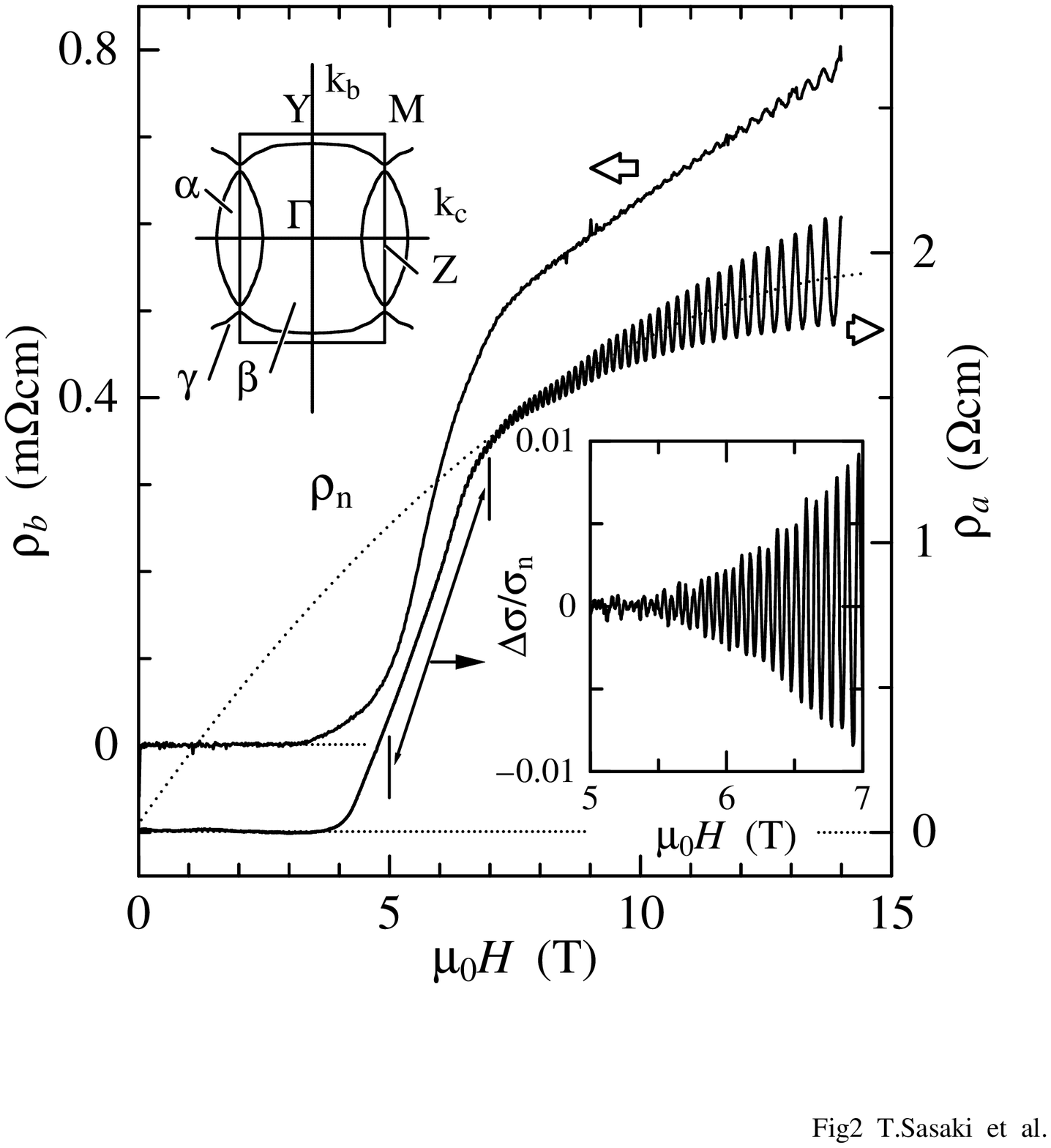}
\caption{Magnetic field dependence of the resistivity $\rho_{b}$ and $\rho_{a}$ along the $b$ (in the Q2D plane) and the $a^{*}$ axes (out of the plane), respectively, in the sample \#2.  The upper left inset shows the first Brillouin zone and the Fermi surface.  The lower right inset indicates the SdH oscillations on $\rho_{a}$ plotted as $\Delta\sigma/\sigma_{\rm n}$.}
\end{figure}

Figure 2 shows the magnetic field dependence of the resistivity in the sample \#2 at $T =$ 0.52 K.  
The in-plane ($\rho_{b}$) and the out of plane ($\rho_{a}$) resistivities are measured in one single crystal along the $b$ and $a^{*}$-axes, respectively.
The magnitude of the magnetic field dependence of the resistivity and the resistivity onset corresponding to $H_{\rm irr}$ are not influenced by the applied current density in both configurations.  
The flux flow resistivity in the QVL region, however, changes a little with the current density because of the weak non-linear resistance.\cite{sasaki2}  
The two curves in this figure are measured using the current density (current) of $J =$ 0.16 A/cm$^{2}$ ($I =$ 100 $\mu$A) for $\rho_{b}$ and 1.6 mA/cm$^{2}$ (10 $\mu$A) for $\rho_{a}$, respectively.  
The SdH oscillations with the frequency of 599 $\pm$ 2~T are clearly observed.  
The oscillations come from the $\alpha$-orbit centered at the Z-point of the first Brillouin zone depicted in the upper left inset.  
In higher magnetic fields, the magnetic breakdown orbit $\beta$, consisting of the $\alpha$ and opened  $\gamma$ orbits, has been observed in both SdH and dHvA oscillations.\cite{sasaki3,meyer}  
We, however, restrict ourselves to the single band model for the following analysis and discussion on the SdH and dHvA effects because the magnetic field used in the present study is smaller than the magnetic breakdown field.\cite{sasaki3,meyer}  
The amplitude of the SdH oscillation on $\rho_{a}$ is much larger than that on $\rho_{b}$, although the magnitude of the magnetic field dependence of the resistivity is almost the same with two configurations. 
The reason is not known at present but the similar tendency is commonly seen in the Q2D organic conductors.\cite{singleton1}
  
In order to see the SdH oscillations in low magnetic field region, $\Delta\sigma/\sigma_{\rm n}$ is shown in the lower right inset of Fig. 2.  
Here, $\Delta\sigma$ is the oscillatory part of the conductivity obtained by subtracting the non-oscillatory part of the conductivity, and $\sigma_{\rm n}$ ($\simeq \rho_{\rm n}^{-1}$) is the normal conductivity which is the same with the non-oscillatory part of the conductivity in the normal state.  
In the vortex state below about 7~T, SdH oscillations come from the normal (quasi particle) component of the total conductivity which includes additional non-equilibrium superconducting component due to the vortex pinning.  
The normal (quasi particle) part of the conductivity roughly corresponds to the flux flow conductivity.
Then the normal resistivity $\rho_{\rm n} (\simeq \sigma_{\rm n}^{-1})$ in the vortex state is assumed to be extrapolated smoothly from the normal state into the vortex state toward $\rho =$ 0 at $H =$ 0. 
The dotted line in Fig. 2 shows the normal (quasi particle) resistivity $\rho_{\rm n} (\simeq \sigma_{\rm n}^{-1})$ in the normal (vortex) state.  
The line is obtained as almost linear in the vortex state and passing through the middle of the SdH oscillations in the normal state. 
In intermediate region around 7~T, two lines are connected smoothly by the simple second order polynomial function. 
The SdH oscillation is persisting down to about 5~T where the resistivity is about 30 \% of the normal state value ($\sim$ 1.5~$\Omega$cm at 7~T).  
In the other sample \#3 the oscillations can be seen in lower magnetic field of about 4.5~T, which are shown latter in Fig. 4.
This is first unambiguous observation of the SdH oscillations on the flux flow resistance in the well characterized superconductor.
Similar SdH effect in the superconducting state has been reported in the $\beta^{\prime\prime}$-type BEDT-TTF organic superconductor.\cite{wosnitza}  
But it did not show the additional amplitude damping in both the SdH and dHvA oscillations in the superconducting state.  
It may be necessary to consider the smaller $H_{\rm c2}$ value reported by another group\cite{mueller2} than the value expected in Ref. \onlinecite{wosnitza}.

\begin{figure}[t]
\includegraphics[viewport=2cm 4cm 20cm 26cm,clip,width=0.9\linewidth]{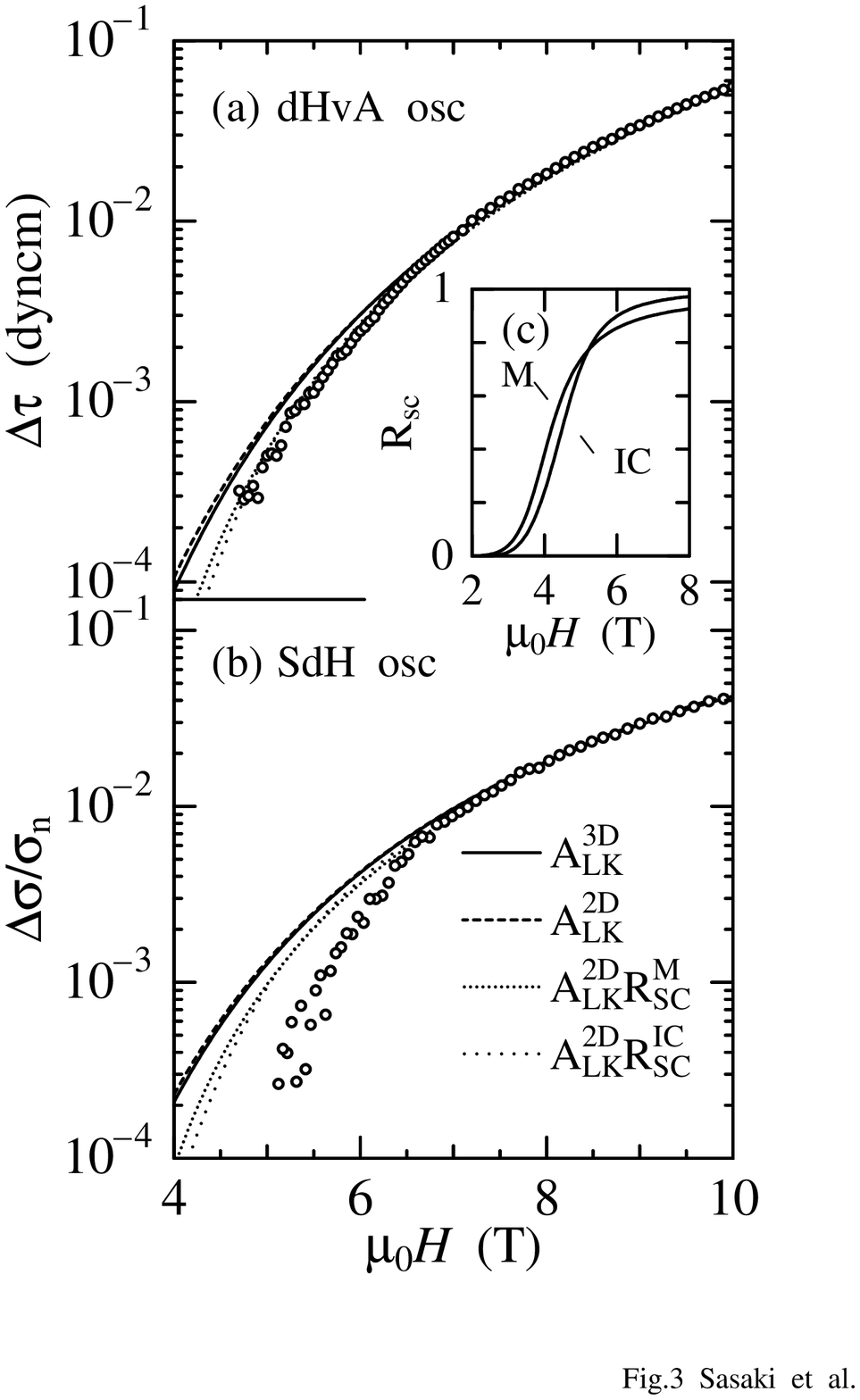}
\caption{Magnetic field dependence of the oscillation amplitude of (a) the dHvA and (b) SdH effects.  (c) The additional damping factor in the vortex state $R_{\rm SC}$; M by Maniv {\it et al.}\cite{maniv1} and IC by Ito {\it et al.}\cite{ito} and Clayton {\it et al.}\cite{clayton}  The solid and broken curves in (a) and (b) are the field dependence of the oscillation amplitude $A_{\rm LK}^{\rm 3D}$ and $A_{\rm LK}^{\rm 2D}$ based on the LK formula for the 3D and 2D cases, respectively.  }
\end{figure}

The additional damping of the SdH and dHvA oscillation amplitude in the vortex state is demonstrated in Fig. 3.  
The oscillation amplitude $\Delta\tau$ and $\Delta\sigma/\sigma_{\rm n}$ for the dHvA and SdH effects is shown by the circles in Figs. 3(a) and 3(b), respectively.  
In the normal state above about 7~T, both oscillation amplitude are well described by the standard Lifshitz-Kosevich (LK) formula.\cite{shoenberg}  
The amplitude $A_{\rm LK}$ of the first harmonics of the oscillations in the single band is given by $A_{\rm LK} \propto TH^{n}R_{T}R_{D}R_{S}$, where the temperature factor $R_{T} = ({\lambda}m_{c}T/H)/\sinh({\lambda}m_{c}T/H)$, the Dingle factor $R_{D} = \exp(-{\lambda}m_{b}T_{D}/H)$, and the spin factor $R_{S} = \cos({\pi}gm_{b}/2m_{0})$.  
Here, $\lambda \equiv 2\pi^{2}ck_{B}/e\hbar =$ 14.69~T/K, $m_{b}$, $m_{c}$ and $m_{0}$ are the band, cyclotron effective and free electron masses, and $T_{D}$ is the Dingle temperature related to the scattering rate $\tau_{0}$ with $T_{D} = \hbar/2{\pi}k_{B}\tau_{0}$.  
The power $n$ of $H$ depends on the measured quantity and the dimensionality.  
For the SdH and torque-dHvA amplitude ($\Delta\sigma/\sigma_{\rm n}$ and $\Delta\tau$) in the 3D (2D) case, $n$ is 3/2 (1).
The fitting in both cases are very good with $m_{c} = 3.5 m_{0}$ and $T_{D} = 0.26 \pm$ 0.05~K $(0.17 \pm 0.03$~K) for dHvA and $0.24 \pm 0.05$~K $(0.16 \pm 0.03$~K) for SdH effects in the 3D (2D) case.\cite{mass}  
The 2D formula, however, is adopted for the latter analysis.  
Because the quantized Landau level spacing in the present magnetic field region is considered to be fairly larger than the interlayer transfer integral of this Q2D organic superconductor.\cite{singleton2}
 
Below about 7T the amplitude starts to deviate smoothly downward from $A_{\rm LK}$ in both dHvA and SdH effects.  
This indicates the additional amplitude damping in the vortex state, which has been reported so far on the dHvA effect in $\kappa$-(BEDT-TTF)$_{2}$Cu(NCS)$_{2}$.\cite{sasaki1,ito,wel,clayton}  
The origin of such smooth damping has been discussed on the basis of the model of the quasiparticle scattering by the random vortex lattice with the large superconducting (vortex) fluctuation around the mean field $H_{\rm c2}$.\cite{maniv1,clayton}  
In the approach by Maniv {\it et al.},\cite{maniv1} the additional damping term $R_{\rm SC}$ as being multiplied to $A_{\rm LK}$ is $R_{\rm SC} = \exp(-\pi^{3/2}{\langle}|\tilde{\Delta}|^{2}{\rangle}/n_{F}^{1/2})$, where $n_{F} \equiv E_{F}/\hbar\omega_{c}$, $\omega_{c} \equiv eH/m_{c}c$, $E_{F}$ the Fermi energy, and ${\langle}|\tilde{\Delta}|^{2}\rangle$ is the mean square of the superconducting order parameter averaged over space coordinates.  
The tilde above $\Delta$ indicates that energy is measured in units of $\hbar\omega_{c}$.  
This expression is very similar to the Maki Stephen type mean field approach.\cite{maki,stephen,clayton}  
In a simple analytic expression derived by Maniv {\it et al.},\cite{maniv1} ${\langle}|\tilde{\Delta}|^{2}\rangle \simeq ({\alpha}/{\beta})[1 + \exp(-x^{2})/2x\int_{-\infty}^{x}\exp(-y^{2})dy]$, where $\alpha = (1/2\hbar\omega_{c})\ln\sqrt{H_{\rm c2}/H}$, $\beta =1.38/n_{F}(\hbar\omega_{c})^{3}$, and $x = \alpha/\sqrt{2{\beta}k_{B}T}$.  
Ito {\it et al.}\cite{ito} and Clayton {\it et al.}\cite{clayton} have used an approximated interpolation formula for the mean square of the gap function, ${\langle}|\Delta|^{2}{\rangle} = \sqrt{[\Delta(0)^{2}(1-H/H_{\rm c2})/2]^{2}+{\alpha}(T)^{2}}+\Delta(0)^{2}(1-H/H_{\rm c2})/2$, where $\alpha(T)$ is a temperature dependent parameter to scale the fluctuations.
Both calculations of $R_{\rm SC}$ are shown in Fig. 3(c) by using the same mean field $H_{\rm c2} =$ 4.8~T.  
The results on $R_{\rm SC}$ are in fairly good agreement with each other.
  
The dotted curves in Figs. 3(a) and 3(b) show the expected magnetic field dependence of the oscillation amplitude taking $R_{\rm SC}$ into account.  
The dHvA effect is well represented in these fluctuation approaches as has been reported.\cite{maniv1,ito,clayton}  
In the SdH effect, the experimental results follow well $A_{\rm LK}^{\rm 2D}R_{\rm SC}$ near $H_{\rm c2} \sim$ 7~T as well as the dHvA effect.  
A stronger damping, however, appears below about 6.5~T in only the SdH effect.  

We discuss the stronger damping in the vortex state observed only in the SdH effect in connection with the quantum vortex slush state at low temperature.
In the QVL region the finite resistance appears in the vortex liquid state between the melting $H_{m}$ or irreversible field $H_{\rm irr}$ and $H_{\rm c2}$ even at $T \sim 0$.  
In the case of less or no quantum fluctuation, the zero resistance should appear just below $H_{\rm c2}$ and $H_{m}(T)$ or $H_{\rm irr}(T)$ coincides with $H_{\rm c2}(T)$ at $T = 0$.  
The former QVL region has been actually found in $\kappa$-(BEDT-TTF)$_{2}$Cu(NCS)$_{2}$ as a demonstration of the importance of the quantum fluctuations in this material.\cite{sasaki1,mola}  
Recently two vortex liquid regions have been found at low temperature.\cite{sasaki2}
The low resistivity state with non-linear current response below about 1~K has been distinguished from the high resistivity state at higher temperature.  
A steep drop of the resistivity around $T_{\rm L} \sim$ 1~K separates the vortex liquid state into these two regions. 
The short-range order of vortices has been expected to exist in the former low resistivity state referred to as the quantum vortex slush state.  
Because these features in the low resistivity state are phenomenologically similar to the observations explained by the the vortex slush concept with the short-range order of vortices in the high-$T_{\rm c}$ oxides.\cite{worthington,nishizaki,nonomura} 
The latter high resistivity state has been considered as the thermal vortex liquid (TVL) state where no translational long-range order of vortices is formed. 

\begin{figure}[t]
\includegraphics[viewport=2cm 4cm 20cm 25cm,clip,width=0.9\linewidth]{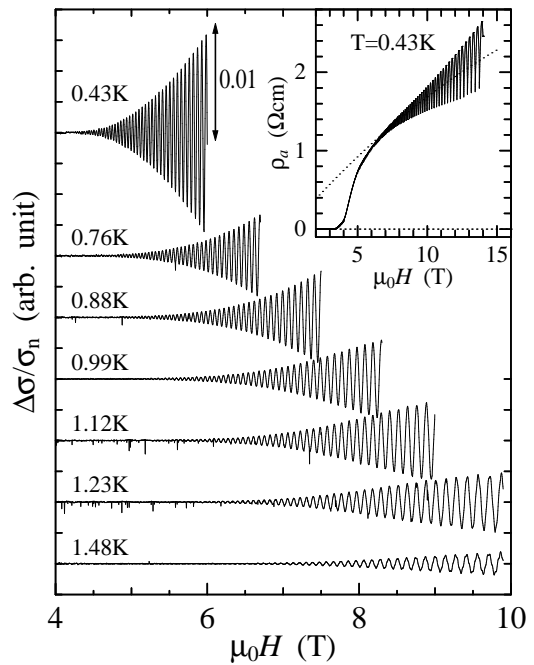}
\caption{SdH oscillations in the sample \#3.  Inset shows the overall magnetic field dependence of the resistivity $\rho_{\rm a}$ at $T =$ 0.43~K.  The dotted line indicates the normal (quasi particle) resistivity $\rho_{\rm n} \simeq \sigma_{\rm n}^{-1}$. }
\end{figure}

Since the additional damping of the oscillation amplitude in the vortex state is explained to be sensitive to the local phase modulation of the superconducting order parameter,\cite{maniv1} some kind of perturbation on the vortices, for example, applying the transport current in the SdH measurements, are expected to influence the phase coherence.  
It is noted that both dHvA and SdH oscillations in this experiment are measured in the quantum vortex slush region.
In the case of such measurements in the TVL state, the additional damping in the vortex state is expected to be the same in dHvA and SdH effects because TVL does not have any order of vortices.  
Thus neither the presence of the transport current in the measurements of the SdH effect nor the absence of current in the dHvA effect alters any phase coherence.  
On the other hand, SdH effect in the quantum vortex slush state may be affected by moving vortices due to applying the transport current.  
It may disturb such coherency of the vortices and quasiparticles.  
Non-observation of stronger oscillation amplitude damping in the dHvA effect with basically no transport current demonstrates the influence of the moving vortices by the current in the quantum vortex slush state.  

Besides applying transport current, the effect of moving vortices on the damping of the dHvA oscillations has been seen in the superconducting state of 2{\it H}-NbSe$_{2}$.\cite{steep}  
The damping of the oscillations depended on the history of reciprocal partial sweeps of the external magnetic field in the hysteretic region where the pinning strength was changing.  
For explaining the observations, Maniv {\it et al.} have suggested that the motion of vortices, depending on the pinning strength, influences the magnitude of the damping of the oscillations.\cite{maniv1}

In order to see the difference of the damping in the quantum vortex slush and TVL regions, we tried to compare the SdH oscillations at higher temperatures.  
Figure 4 shows the SdH oscillations at higher temperatures in the sample \#3.  
The SdH oscillations in the vortex state are clearly seen down to about 4.5~T at 0.43~K.  
The sample \#3 seems to have better quality to see the SdH oscillations in lower magnetic field than that in the sample \#2 presented in Fig. 1.
\begin{figure}[t]
\includegraphics[viewport=2cm 5cm 21cm 26cm,clip,width=0.9\linewidth]{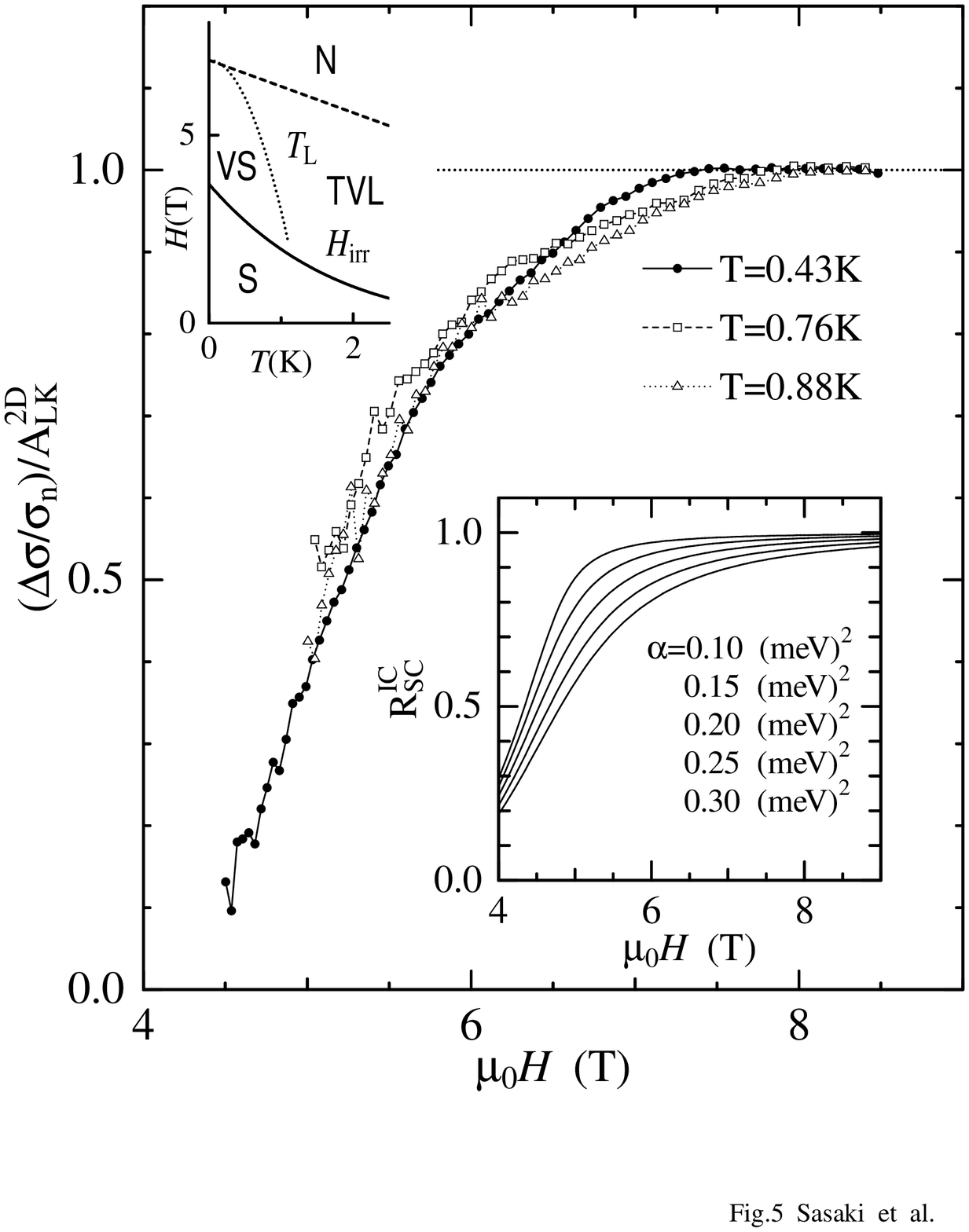}
\caption{SdH oscillation amplitude scaled by $A_{\rm LK}^{\rm 2D}$ in the sample\#3 at 0.43, 0.76, and 0.88~K.  Upper left inset shows the low temperature part of the schematic vortex phase diagram.\cite{sasaki2}  N is the normal, S the vortex solid, VS and TVL the quantum vortex slush and thermal vortex liquid states, respectively.  Lower inset demonstrates the additional damping factor proposed by Ito {\it et al.}\cite{ito} and Clayton {\it et al.}\cite{clayton}  Each curve from top to bottom is calculated with $\alpha =$ 0.10, 0.15, 0.20, 0.25, and 0.30 (meV)$^{2}$. }
\end{figure}

The magnetic field dependence of the observed SdH oscillation amplitude scaled by $A_{\rm LK}^{\rm 2D}$ in the normal state is shown in Fig. 5.  
The value of $(\Delta\sigma/\sigma_{\rm n})/A_{\rm LK}^{\rm 2D}$ corresponds to the additional damping in the vortex state.  
The additional damping starts smoothly around $H_{\rm c2} \simeq$ 7~T.  
The magnitude of the additional damping becomes larger in higher temperatures near $H_{\rm c2}$.  
It is naturally understood by the fluctuations.  
The lower inset shows calculations of the additional damping factor $R_{\rm SC}$ proposed by Ito {\it et al.}\cite{ito} and Clayton {\it et al.}\cite{clayton}
The fluctuation parameter $\alpha$ is expected to take larger value in higher temperature.  
This has been actually confirmed in the dHvA experiments on the same material by Clayton {\it et al.}\cite{clayton}
The $\alpha$ value has changed almost continuously from 0.13 at 0.03~K to 0.32~(meV)$^{2}$ at 0.44~K. 
It is noted that the $R_{\rm SC}$ curves calculated with $\alpha$ or corresponding temperature are continuously shifting to smaller value, and do not cross each other.  
The present SdH oscillation damping near $H_{\rm c2}$ is in agreement with the theoretical calculations.  
But the oscillation amplitude at 0.43~K shows stronger additional damping in low magnetic field than others measured at high temperature.  
This stronger damping observed at 0.43~K is considered to be due to the moving vortices in the quantum vortex slush state as discussed above.  

It is expected that the similar stronger damping may appear below about 5.5, and 5 T at 0.76 and 0.88~K, respectively, where the boundary between the quantum vortex slush and the thermal vortex liquid regions is located.\cite{sasaki2}  
This boundary can be seen in the schematic vortex phase diagram in the low temperature region\cite{sasaki2} shown in the upper left inset of Fig. 5.  
In the diagram, N is the normal, S the vortex solid, VS and TVL the quantum vortex slush and thermal vortex liquid states, respectively.  
The SdH oscillation amplitude at 0.76 and 0.88~K, however, does not show the clear stronger damping down to the lowest magnetic field where the SdH oscillations can be detected at those temperatures.  
It means that the magnetic fields where the SdH oscillations are observed are mostly still in the TVL region.  
Then the magnetic field dependence of the damping seems to follow the damping by the thermal fluctuations with temperature.  
This result suggests that the expected stronger damping may appear at lower magnetic fields.

In order to confirm the proposed model on the stronger damping in the SdH oscillations in the quantum vortex slush state, it is necessary to show the clear relation between the stronger damping and the vortex phase diagram.  
More precise measurements at lower magnetic fields in better quality sample are required in future.  
The transport current density dependence of the SdH oscillation amplitude is also important to measure in the quantum vortex slush state because the non-linear behavior in the resistivity has been observed.  
These experiments are in progress.  

\section{Summary}

We observed the SdH oscillations on the flux flow resistivity in the vortex state of the Q2D organic superconductor $\kappa$-(BEDT-TTF)$_{2}$Cu(NCS)$_{2}$.  
The additional damping of the SdH oscillation amplitude near $H_{\rm c2}$ is well described by the model on the superconducting (vortex) fluctuations as well as those observed in the dHvA effect.
In only the SdH effect, the stronger damping appears in the quantum vortex slush region.
The stronger damping may reflect the perturbation of a phase coherence of vortices and quasiparticles in the quantum vortex slush state with the short-range order of vortices due to moving those by the transport current.  

\begin{acknowledgments}
The authors thank K. Kishigi, T. Maniv, T. Nishizaki for stimulating discussions.  
A part of this work was performed at the High Field Laboratory for Superconducting Materials, IMR, Tohoku University.  
This work was partly supported by a Grant-in-Aid for Scientific Research from the Ministry of Education, Science, Sports, and Culture of Japan.

\end{acknowledgments}


\end{document}